\documentclass[prl,twocolumn]{revtex4}
\usepackage{graphicx}

\newcommand{\Wcm}{\,\mathrm{W/cm}^2}
\newcommand{\Up}{U_{\rm P}}

\newcommand{\E}{\mathcal{E}}

\newcommand{\PRL}{{Phys. Rev. Lett.\ }}
\newcommand{\PRA}{{Phys. Rev. A\ }}

\begin{document}
\title{Attosecond double-slit experiment}
 
\author{F. Lindner,$^{1}$ M. G. Sch\"{a}tzel,$^{1}$ H. Walther,$^{1,2}$ 
A. Baltu\v{s}ka,$^{1}$ E. Goulielmakis,$^{1}$ F. Krausz,$^{1,2,3}$
D. B. Milo\v{s}evi\'{c},$^{4}$ D. Bauer,$^{5}$ W. Becker,$^{6}$ and G. G. Paulus$^{1,2,7}$ } 
\affiliation{$^1$Max-Planck-Institut f\"{u}r Quantenoptik, 85748 Garching, Germany}
\affiliation{$^2$Ludwig-Maximilians-Universit\"{a}t M\"{u}nchen, 85748 Garching, Germany}
\affiliation{$^3$Institut f\"{u}r Photonik, Technische Universit\"{a}t Wien, Gusshausstr. 27, A-1040 Wien, Austria}
\affiliation{$^4$Faculty of Science, University of Sarajevo, Zmaja od Bosne 35, 71000 Sarajevo, Bosnia and Hercegovina}
\affiliation{$^5$Max-Planck-Institut f\"{u}r Kernphysik, Saupfercheckweg 1, 69117 Heidelberg, Germany}
\affiliation{$^6$Max-Born-Institut, Max-Born-Str.  2a, 12489 Berlin, Germany}
\affiliation{$^7$Department of Physics, Texas A\&M University, College Station, TX 77843-4242}

\date{\today}
%\draft
\begin{abstract}
A new scheme for a double-slit experiment in the time domain is presented. Phase-stabilized few-cycle laser pulses open one to two windows (``slits'') of attosecond duration for photoionization. Fringes in the angle-resolved energy spectrum of varying visibility depending on the degree of which-way information are observed. A situation in which one and the same electron encounters a single and a double slit at the same time is discussed. The investigation of the fringes makes possible interferometry on the attosecond time scale. The number of visible fringes, for example, indicates that the slits are extended over about 500\,as.  

\vspace{2cm}

\end{abstract}
\pacs{ 03.65.Ta, 07.60.Ly, 32.80.Rm}

\maketitle

The conceptually most important interference experiment is the double-slit scheme, which has played a pivotal role in the development of optics and quantum mechanics. In optics its history goes back to Young's double-slit experiment. Its scope was greatly expanded by Zernike's work and continues to deliver new insights into coherence to the present day \cite{MandelWolf}. One of the key postulates of quantum theory is interference of matter waves, experimentally confirmed by electron diffraction \cite{Davisson,Thomson}. More than 30 years later, J\"{o}nsson was the first to perform a double-slit experiment with electrons \cite{Joensson}. Of particular importance for interpreting quantum mechanics have been experiments with a single particle at any given time in the apparatus \cite{Merli,Tonomura}. More recent work has illuminated the fundamental importance of complementarity in which-way experiments \cite{ScullyNature91} and of quantum information in quantum-eraser schemes \cite{ShihPRL00}.

In this letter a novel realization of the double-slit experiment is described. It is distinguished from conventional schemes by a combination of characteristics: {\it (i)} The double slit is realized not in position-momentum but in time-energy domain. {\it (ii)}  The role of the slits is played by windows in time of attosecond duration. {\it (iii)} These ``slits'' can be opened or closed by changing the temporal evolution of the field of a few-cycle laser pulse. {\it (iv)} At any given time there is only a single electron in the double-slit arrangement. {\it (v)} The presence and absence of interference are observed for the same electron at the same time.

\begin{figure}
\includegraphics[width=0.9\columnwidth]{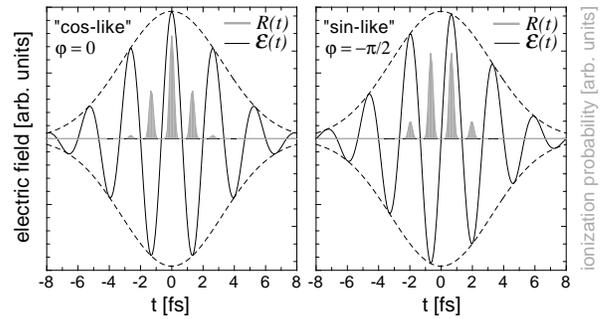}
\caption{\label{fig1}Temporal variation of the electric field $\E(t)=\E_0(t)\cos(\omega t+\varphi)$ of few-cycle laser pulses with phase $\varphi=0$ (``cosine-like'') and $\varphi=-\pi/2$ (``sine-like''). In addition, the field ionization probability $R(t)$, calculated at the experimental parameters, is indicated. Note that an electron ionized at $t=t_0$ will not necessarily be detected in the opposite direction of the field $\E$ at time $t_0$ due to deflection in the oscillating field.  }
\end{figure}

Interference experiments in the time-energy domain are not entirely new. Interfering electron wave packets were created by femtosecond laser pulses \cite{Wollenhaupt02}. Accordingly, the windows in time (or temporal slits) during which these wave packets are launched were comparable to the pulse duration. In the present experiment, in contrast, the slits are open during a small fraction of an optical cycle, which gives the attosecond width. A number of experiments, in particular in intense-laser atom physics, can and have been interpreted in this spirit (for a review see, for example, \cite{BeckerAdvances02}), and were also extended to the microwave region \cite{gallagher88}. Here, however, the optical cycles are precisely tailored by controlling the phase of few-cycle laser pulses (also known as absolute or carrier-envelope phase). This provides an unprecedented degree of control for the double-slit arrangement. Not only are the principles of quantum mechanics beautifully demonstrated, it is also likely that applications exploiting interferometric techniques for measuring attosecond dynamics will emerge.

Argon atoms are ionized by intense few-cycle 850-nm laser pulses. Photoionization under these conditions is a highly nonlinear process whose first step  can be described by optical field ionization. This immediately explains the generation of one attosecond window (or slit) in time per half-cycle close to its extremum, see Fig. 1. By using phase-controlled few-cycle laser pulses \cite{BaltuskaNature03}, it is possible to manipulate the temporal evolution of the field, thus gradually opening or closing the slits, and controlling which-way information. Depending on the field, one or two half-cycles (or anything in between) contribute to the electron amplitude for a given direction and electron energy. This corresponds to a varying degree of which-way information and, accordingly, to varying contrast of the interference fringes. Subsequent half-cycles emit electrons in opposite directions. The temporal slits are therefore spaced by approximately the optical period, resulting in a fringe spacing close to the photon energy. 

\begin{figure}
\includegraphics[width=0.95\columnwidth]{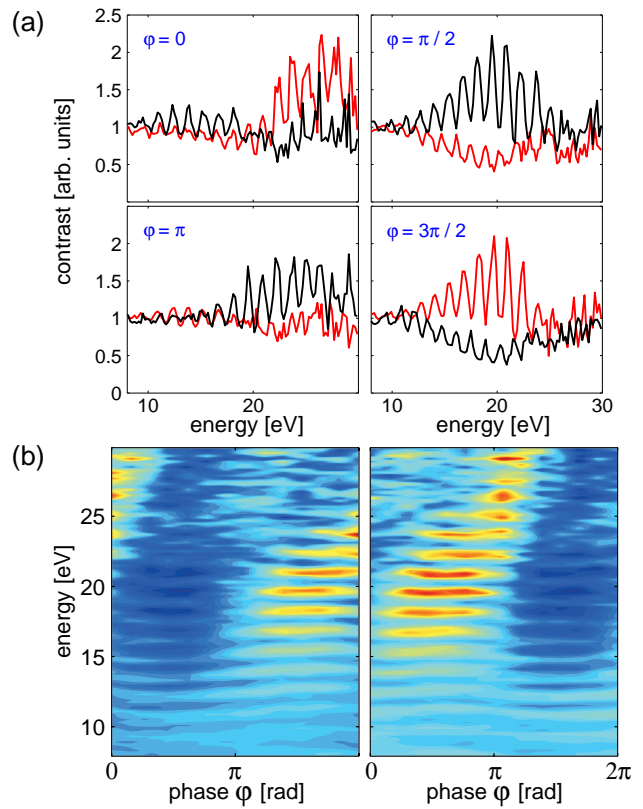}
\caption{\label{fig2} Photoelectron spectra of argon measured with 6-fs laser pulses for  intensity $1\times 10^{14}\Wcm$ as a function of the phase. Panel (a) displays the spectra for $\pm$sine- and $\pm$cosine-like laser fields. The red curves are spectra recorded with the left detector (negative direction), while the black curves relate to the positive direction. For $\varphi=\pi/2$ the fringe exhibit maximum visibility for electron emission to the right, while in the opposite direction minimum fringe visibility is observed. In addition, the fringe positions are shifted. Panel (b) displays the entire measurement where the fringe visibility is coded in false colors. The fringe positions vary as the phase $\varphi$ of the pulse is changed. This causes the wave-like bending of the stripes in these figures. Both panels, in principle, show redundant information because a phase shift of $\pi$ mirrors the pulse field in space and thus reverses the role of positive and negative direction. However, this data was in fact measured simultaneously and thus single- and double-slit behavior is observed for the same electron at the same time.}
\end{figure}

The experimental setup is quite similar to that described in \cite{PaulusPRL03}. The laser beam mentioned above intersects an atomic gas jet inside a vacuum apparatus. The laser polarization is horizontal and electrons emitted in opposite directions (``left'' and ``right'') are detected by two opposing time-of-flight (TOF) detectors. The phase of the field can be controlled by delaying the envelope of the pulse with respect to the carrier by means of a glass wedge shifted into or out of the beam. The phase of the field is measured as described in \cite{PaulusPRL03}.

Figure 2 displays measured electron spectra. In Fig.~2(a) the spectra recorded at the left and the right detectors are shown for $\pm$cosine-like and $\pm$sine-like pulses as defined in Fig. 1. A problem in presenting such spectra is that they quickly roll off with increasing electron energy. This roll-off was eliminated by dividing the spectra by the average of all spectra over the pulse's phase. Clear interference fringes with varying visibility are observed as expected from the discussion above. The highest visibility is observed for $-$sine-like pulses in the positive (``right'') direction. For the same pulses, the visibility is very low in the opposite direction. Changing the phase by $\pi$ reverses the role of left and right as expected. The most straightforward explanation -- which will be detailed by a simple model below -- is to assume that, for $-$sine-like pulses, there are two slits and no which-way information for the positive direction and just one slit and (almost) complete which-way information in the negative direction. The fact that the interference pattern does not entirely disappear is caused by the pulse duration, which is still slightly too long to create a perfect single slit. It should be noted at this point that there is only a single photoelectron involved at a time because single ionization is observed. At the same time, this single electron interferes in one direction and does not in the other. 

The fringe pattern exhibits an envelope. From Fig. 2 a width of this envelope of about 4 fringes is inferred. Just as for a double-slit experiment, the width of this envelope can be associated with the width of the slits. It will turn out, however, that what is seen here is not the width of the slit. Rather, each slit can be resolved into a pair of slits whose separation is inversely proportional to the width of the envelope.

Disregarding the changing visibility, the peaks observed in the spectra resemble the well-known above-threshold ionization (ATI) peak pattern and they are certainly related to them. However, the relationship is non-trivial: Besides the visibility of the fringes, also their positions change as the phase of the field is varied. Details of the fringe shifts can be seen in Fig.~2(b). For conventional ATI, one would try to explain this in terms of the ponderomotive potential $\Up$. This does not work here, because the concept of the ponderomotive potential, which is defined as the cycle-averaged kinetic energy of an electron quivering in an oscillating electric field, is questionable in the few-cycle regime.  

In contrast, an interpretation based on the double-slit analogy is obvious. In a spatial double slit, the fringe pattern would shift if a phase shifter (for light, simply a glass plate) were placed in front of one of the slits. For nontrivial particle trajectories one needs to consider the action $S$ along the particles' paths and use the fact that the particles' phases are given by $S/\hbar$. 

\begin{figure}
\includegraphics[width=0.55\columnwidth]{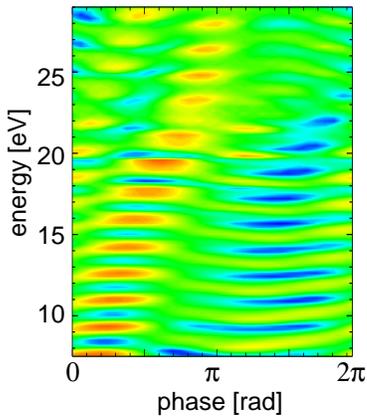}
\caption{\label{fig3}The result of a numerical solution of the time-dependent Schr\"{o}dinger equation, as described in the text. This figure should be compared with the right panel of Fig. 2(b).}
\end{figure}

In order to exclude other scenarios, we compare the experimental data with results obtained by numerically solving the time-dependent Schr\"{o}dinger equation (TDSE) in three spatial dimensions. The self-consistent effective argon potential was calculated numerically within density-functional theory using the optimized effective potential approach proposed in Ref.\ \cite{kli}. During time-propagation, only the $3p$ valence electron was considered active, moving in the combined field of the effective potential and the $6.5$-cycle $\sin^2$ laser pulse of peak intensity $10^{14}$\,W/cm$^2$. The directional spectra were calculated for 32 different carrier-envelope phases using a method 
described in \cite{bauerPRL}. Finally, the spectra were divided by the phase-averaged spectrum, using the same procedure that was applied to the experimental data underlying  Fig.\,\ref{fig2}. The numerical TDSE result for the right-going electrons is shown in Fig. 3, to be compared with the experimental result in the right panel of Fig. 2(b). Virtually all details found in the measurement can also be found in the calculation. This confirms that single-electron dynamics are sufficient to explain the fringes.

For an interpretation we resort to a classical model, the so-called simple-man's model \cite{Longname88}, which -- together with various extensions and modifications -- has proven to be extremely helpful for understanding strong-field laser-atom interaction; for a review see, for example, \cite{BeckerAdvances02}. Alternatively, Keldysh-type models, which can be interpreted as an approximation of Feynman's path integral \cite{SalieresScience01}, could be used. Respective results can be found in the literature: Ref.~\cite{MilosevicPRL02} predicts effects analogous to those described in this letter for circular polarization. References \cite{BaltuskaNature03,NisoliPRL03,SansonePRL04} explain related classical effects for electromagnetic XUV radiation produced by high-harmonic generation. For the present problem, the classical and the quantum model lead to qualitatively the same results.

The classical model assumes that an electron is launched into the continuum at some time $t_0$. Evidently, only for times $t_0$ where the electric field is close to its highest strength, is there an appreciable probability of such a process. Another crucial assumption of the model is that the electron's velocity is zero at $t=t_0$. This means that $p-eA(t_0)=0$, where $p$ is the momentum of the electron at the detector, $A(t)$ the vector potential of the field, and $e=-|e|$ the electron's charge. It is largely this relationship that explains the double-slit behavior of few-cycle photoionization.

\begin{figure}
\includegraphics[width=0.9\columnwidth]{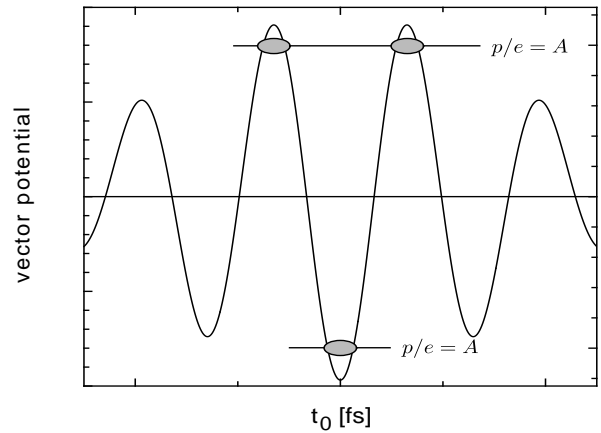}
\caption{\label{fig4}Vector potential of a $-$sine-like few-cycle pulse. The temporal slits are given by the condition $p-eA(t_0)=0$. For a $-$sine-like pulse, this leads to a double slit in the negative direction (since $e=-|e|$) and a single slit in the opposite direction. Each slit can be resolved into a pair of slits. }
\end{figure}

The strength of the classical model is the intuitive insight it provides. In the following, hardly more than the number and position of the solutions of $p-eA(t_0)=0$ for given $p$ will be used in order to explain the double-slit behavior. The respective solutions $t_0(p)$ in a Keldysh-type model are complex, thus allowing  access to classically forbidden electron energies. However, the symmetry of these solutions stays the same and so do the results qualitatively. 

In Fig. 4 the vector potential $A(t)$ is drawn for a $-$sine-like pulse. The solutions of $p-eA(t_0)=0$ and thus all trajectories of momentum $p$ that could interfere can be found by intersecting $A(t)$ with a horizontal line at $p/e$. It is now important to recall that a fringe pattern of maximal visibility requires equally ``strong'' slits, i.e. minimal which-way information. For a few-cycle pulse whose envelope is maximal at $t=0$, the ``strength'' of a slit decreases very quickly with increasing $|t_0|$  and is essentially zero for $|t_0|>2\pi/\omega$ because of the highly nonlinear dependence of photoionization on the field strength. As the maximum of the pulse envelope was chosen to be at $t=0$, the condition of equally strong slits is identical to requiring that the solutions of $p-eA(t_0)=0$ be symmetric with respect to $t=0$. This is the case for $-$sine-like pulses with electrons emitted in the negative direction and for $+$sine-like pulses with electrons emitted in the positive direction. For both cases, the respective opposite direction can be considered to act like a single slit as long as the pulse is short enough. 

Figure 4 also shows that each slit is, on closer inspection, a pair of slits and that the temporal separation of these sub-slits depends on the electron energy \cite{footnote}. The experimental data must be considered to be a measurement of the time difference of the two sub-slits, which is approximately 500\,as. This is a first simple example for using  interferometry on the attosecond time scale in order to investigate electronic dynamics. In addition, Fig. 2(a) shows that the relative phase of the sub-slits is different for sine- and cosine-like pulses, resulting in a shift of the fringe envelope.
 
It should be noted that the simple-man's model does not reproduce the dependence of the fringe visibility on electron energy as observed experimentally and in the solution of the TDSE. Therefore, the direction for which interference is predicted by the simple model may be wrong, depending on the energy. Using several theoretical models (3D TDSE, 1D TDSE, Keldysh-type, and classical), we were able to show that this is not a fundamental problem of the classical theory. Rather, it is an effect of the atomic binding potential, which obviously deflects the outgoing photoelectrons. The solution of the \textit{one-dimensional} TDSE (which cannot deflect) with a soft-core potential, for example, agrees qualitatively very well with the classical and a Keldysh-type model. In particular, it does not show a pronounced energy dependence of the fringe visibility, and it predicts the interferences in the same direction as the simple models.

More insight from the classical model can be gained by treating the electrons as deBroglie-waves, computing their actions $S$, calculating their phases $S/\hbar$, and adding them coherently. This allows predicting the fringe positions: Just as for any other double slit, fringe maxima are observed, if the difference in phase is $n\cdot 2\pi$ were $n$ is the order of the fringe. Indeed, respective calculations show a phase-dependent fringe shift resembling the one experimentally observed. In the same way, also maxima and minima of the fringe pattern's envelope can be calculated in dependence of the phase $\varphi$. However, quantitative agreement is certainly not to be expected, given that the classical model neglects the atomic potential entirely.

In conclusion, we have realized an intriguing implementation of the double slit in the time domain. The observation of interference and its absence at the same time for the same electron is a beautiful demonstration of the principles of quantum mechanics. It should also be noted that attosecond slits were used and that the interferograms reflect the attosecond dynamics of electronic transitions. Further experimental and theoretical progress should make it possible to use interferometric techniques for attosecond science.

This work has been supported by the Austrian Science Fund (Grants No.~F016, No.~Z63, and No. P15382), the German Research Foundation (Grant No.~PA730/2 and BA2190/1), the Welch Foundation (Grant No. A-1562), and VolkswagenStiftung.


\begin{references}
\bibitem{footnote} The time separation has to be calculated from the Keldysh-type theory. Note that the classical model of Fig.~4 does not apply for momenta larger than $|e\max A(t)|$
\end{references}
\end{document}